\documentclass[aps,prd,preprintnumbers,showpacs,showkeys,nofootinbib,
superscriptaddress,fleqn,floatfix,tightenlines,10pt]{revtex4-1}
\usepackage{amsmath,amsfonts,amssymb,amscd,amsxtra,amsthm}
\usepackage{graphicx}  
\usepackage{epstopdf}
\usepackage{dcolumn}  
\usepackage{bm}          
\usepackage{slashed}
\usepackage{cancel}
\usepackage{float} 
\usepackage{mathtools}
\usepackage{amsbsy}
\usepackage{amstext}

\providecommand{\tabularnewline}{\\}
\usepackage[utf8]{inputenc} 
\usepackage{booktabs}
\usepackage[normalem]{ulem} 
\usepackage[dvipsnames]{xcolor} 
\usepackage{tabularx}
\usepackage{enumitem}  
\usepackage{array} 
\usepackage{slashed}
\usepackage{tikz}
\usepackage{cleveref} 
\usepackage{multirow}
\renewcommand\sout{\bgroup \color{red} \ULdepth=-.5ex \ULset}

\makeatletter

\begin{document}  
\preprint{INHA-NTG-02/2018}
\title{Magnetic moments of the lowest-lying singly heavy baryons} 
\author{Ghil-Seok Yang}
\email[E-mail: ]{ghsyang@ssu.ac.kr}
\affiliation{Department of Physics, Soongsil University, Seoul 06978,
Republic of Korea}
\author{Hyun-Chul Kim}
\email[E-mail: ]{hchkim@inha.ac.kr}
\affiliation{Department of Physics, Inha University, Incheon 22212,
Republic of Korea}
\affiliation{Research Center for Nuclear Physics (RCNP), Osaka
University, Ibaraki, Osaka 567-0047, Japan}
\affiliation{School of Physics, Korea Institute for Advanced Study 
  (KIAS), Seoul 02455, Republic of Korea}
\date{\today}
\begin{abstract}
A light baryon is viewed as $N_c$ valence quarks bound by meson
mean fields in the large $N_c$ limit. In much the same way a singly
heavy baryon is regarded as $N_c-1$ valence quarks bound by the same
mean fields, which makes it possible to use the properties of light
baryons to investigate those of the heavy baryons. A heavy quark being 
regarded as a static color source in the limit of the infinitely heavy
quark mass, the magnetic moments of the heavy baryon are determined
entirely by the chiral soliton consisting of a light-quark pair. The
magnetic moments of the baryon sextet are obtained by using the
parameters fixed in the light-baryon sector. In this mean-field
approach, the numerical results of the magnetic moments of the baryon
sextet with spin $3/2$ are just 3/2 larger than those with spin
$1/2$. The magnetic moments of the bottom baryons are the same as
those of the corresponding charmed baryons.  
\end{abstract}
\maketitle
\section{Introduction} 
Very recently, Ref.~\cite{Yang:2016qdz} showed that when the number of
colors ($N_c$) goes to infinity singly heavy baryons can be described
as $N_c-1$ valence quarks bound by the meson mean fields that also
have portrayed light baryons as $N_c$ valence quarks bound by the same
mean fields~\cite{Witten:1979kh, Witten:1983tx}, being motivated by
Diakonov~\cite{Diakonov:2010tf}. The masses of the lowest-lying singly
heavy baryons were well reproduced in both the charmed and bottom
sectors, and the mass of the $\Omega_b$ 
was predicted within this framework. Using the method developed in
Ref.~\cite{Yang:2016qdz}, we were able to interpret two narrow
$\Omega_c$ resonances as exotic baryons belonging to the
anti-decapentaplet ($\overline{\bm{15}}$)~\cite{Kim:2017jpx} among five
$\Omega_c$s found by the LHCb Collaboration~\cite{Aaij:2017nav}. 
This mean-field approach is called the chiral quark-soliton model 
($\chi$QSM)~\cite{Diakonov:1987ty} (for a review, see 
Refs.~\cite{Christov:1995vm, Diakonov:1997sj} and references 
therein).  Very recently, the model has also described successfully
strong decays of heavy baryons~\cite{Kim:2017khv} including those of
the newly found two narrow $\Omega_c$s. 

The magnetic moments of the heavy baryons have been already investigated
within various different approaches such as quark
models~\cite{Johnson:1976mv, Choudhury, Lichtenberg:1976fi}, the MIT
bag model~\cite{Bose:1980vy}, the quark potential model~\cite{Jena:1986xs,
  Glozman:1995xy}, the Skyrme models in bound-state
approaches~\cite{Oh:1991ws, Scholl:2003ip}, a 
relativistic  quark model~\cite{Faessler:2006ft}, lattice
QCD~\cite{Can:2013tna, Can:2015exa, Bahtiyar:2016dom}, heavy-baryon
chiral perturbation theories~\cite{Savage:1994zw, Banuls:1999mu}, and
QCD sum rules~\cite{Aliev:2001ig, Zhu:1997as}, and so on. Since there
are no experimental data available yet, it is of great interest to
compare the results with those from several different
approaches. Since the baryon anti-triplet ($\overline{\bm{3}}$) 
consists of the light-quark pair with the total light-quark spin
$J=0$, the corresponding magnetic moment vanishes in the present
mean-field approach with the infinitely heavy-quark mass limit
considered.  Thus, in the present work, we want to employ the
$\chi$QSM to compute the magnetic moments of the lowest-lying singly
heavy baryons, in particular, the baryon sextet ($\bm{6}$) with both
spin $J'=1/2$ and $J'=3/2$.   The magnetic moments of the light
baryons were already studied within the $\chi$QSM~\cite{Kim:1995ha,
  Kim:1997ip}. A merit of this approach is that we can deal with
light and heavy baryons on the same footing. All the dynamical
parameters required for the present analysis were determined in
Ref.~\cite{Yang:2015era} based on the experimental data on the
magnetic moments of the baryon octet, we have no additional free
parameter to handle for those of the heavy baryons.  We obtain the
results for the magnetic moments of the baryon sextet and compare them
with those from other models and lattice QCD. The results turn out to
be consistent with those from the other works, in particular, with
those from Ref.~\cite{Faessler:2006ft}. Compared with the results from
the lattice QCD~\cite{Can:2013tna, Can:2015exa, Bahtiyar:2016dom}, the
present ones are consistently larger than them except for the
$\Sigma_c^{++}$ magnetic moment. 

The structure of the present work is sketched as follows: In Section
II, we briefly review the general formalism of the $\chi$QSM in order
to compute the magnetic moments of the heavy baryons. In
Section III, we show how to carry out the calculation of the magnetic
moments within the present framework, using the dynamical parameters
fixed in the light baryon sector.  
In Section IV, we present the numerical results of the magnetic
moments of the heavy baryons, examining the effects of
flavor $\mathrm{SU(3)}_{\mathrm{f}}$ breaking. We summarize the
present work in the last Section.
\section{General formalism}
In the mean-field approach, a heavy baryon can be expressed by the
correlation function of the $N_c-1$ light-quark operators,  while 
a heavy quark inside it is regarded as a static color source in the
limit of the infinitely heavy quark mass ($m_Q\to \infty$). The heavy
quark is required only to makes the heavy baryon a color singlet
state. The electromagnetic current we now consider consists 
of both the light and heavy quark currents
\begin{align}
  \label{eq:LHcurrent}
J_\mu (x) = \bar{\psi} (x) \gamma_\mu \hat{\mathcal{Q}} \psi(x) + e_{Q}
  \bar{Q}   \gamma_\mu Q,  
\end{align}
where $\hat{\mathcal{Q}}$ denotes the charge operator of the light
quarks in flavor SU(3) space, defined by
\begin{align}
 \label{eq:chargeOp}
\hat{\mathcal{Q}} =
  \begin{pmatrix}
   \frac23 & 0 & 0 \\ 0 & -\frac13 & 0 \\ 0 & 0 & -\frac13
  \end{pmatrix} = \frac12\left(\lambda_3 + \frac1{\sqrt{3}} \lambda_8\right).
\end{align}
Here, $\lambda_3$ and $\lambda_8$ are the well-known flavor SU(3)
Gell-Mann matrices. The $e_Q$ in the second part of the
electromagnetic current in Eq.~(\ref{eq:LHcurrent}) stands for the
heavy-quark charge, which is given as $e_c=2/3$ for the charm quark or
as $e_b=-1/3$ for the bottom quark. The magnetic moment of a
heavy quark is proportional to the inverse of the corresponding mass, 
i.e. $\bm{\mu} \sim (e_Q/m_Q) \bm{\sigma}$, so that it should be very
small in comparison with the light-quark contributions. It plays an essential role only
in describing the baryon anti-triplet, which is understandable,
because the light-quark pair constitutes a spin-zero state. However,
its effect is rather small when it comes to the baryon sextet. 
In the lattice QCD~\cite{Can:2013tna, Can:2015exa, Bahtiyar:2016dom},
it is known that the contribution of the heavy quark to the
magnetic moments of the baryon sextet is approximately one order
smaller than the light-quark
contributions. Ref.~\cite{Faessler:2006ft} also examined the
heavy-quark contribution separately and found that its effect is in
general tiny on the magnetic moments of the baryon sextet.

In principle, one could consider the heavy-quark effects on the
magnetic moments as done in the quark models.  It would give an
overall constant contribution to the magnetic moments of the baryon
sextet such as 
$-e_Q/6m_Q$~\cite{Faessler:2006ft}, which is parametrically  
very small. However, if one wants to consider the heavy-quark
contribution within the present formalism consistently, one should go
beyond the mean-field approximation. This is yet a difficult task,
since we do not know proper nonperturbative interactions between the
light and heavy quarks. Thus, we want to restrict ourselves to the
light-quark contribution from the mean-field approximation, so we will
ignore in the present work that from the heavy quark current in the
limit of $m_Q\to \infty$.  

Hence, we will deal with the first term of
Eq.~\eqref{eq:LHcurrent} when we compute the magnetic moments of heavy
baryons by considering the following baryon matrix elements:    
\begin{align}
\langle B_Q | \bar{\psi}  (x) \gamma_\mu \hat{\mathcal{Q}} \psi(x)
  |B_Q\rangle.   
\label{eq:MatrixEl1}
\end{align}
Since we have ignored the heavy-quark contributions, we obtain the
same results for both the charmed and bottom baryons. So, we will
mainly focus on the magnetic moments of the charmed baryon
sextet in the present work.

The general expressions for the magnetic moments of light baryons have
been constructed already in previous works~\cite{Kim:1995mr,
  Kim:1995ha, Kim:1997ip, Yang:2015era}. We will extend the formalism
for those of heavy baryons in this work. Taking into account the
rotational $1/N_c$ and linear $m_{\mathrm{s}}$ corrections, we are able to write
the collective operator for the magnetic moments as 
\begin{align}
  \label{eq:MagMomOp}
 \hat{\mu} = \hat{\mu}^{(0)}  + \hat{\mu}^{(1)}, 
\end{align}
where $\hat{\mu}^{(0)}$ and $\hat{\mu}^{(1)}$ represent the leading
and rotational $1/N_c$ contributions, and the linear $m_{\mathrm{s}}$
corrections respectively 
\begin{align}
\hat{\mu}^{(0)} & =  
\;\;w_{1}D_{\mathcal{Q}3}^{(8)}
\;+\;w_{2}d_{pq3}D_{\mathcal{Q}p}^{(8)}\cdot\hat{J}_{q}
\;+\;\frac{w_{3}}{\sqrt{3}}D_{\mathcal{Q}8}^{(8)}\hat{J}_{3},\cr
\hat{\mu}^{(1)} & =  
\;\;\frac{w_{4}}{\sqrt{3}}d_{pq3}D_{\mathcal{Q}p}^{(8)}D_{8q}^{(8)}
+w_{5}\left(D_{\mathcal{Q}3}^{(8)}D_{88}^{(8)}+D_{\mathcal{Q}8}^{(8)}D_{83}^{(8)}\right)
\;+\;w_{6}\left(D_{\mathcal{Q}3}^{(8)}D_{88}^{(8)}-D_{\mathcal{Q}8}^{(8)}D_{83}^{(8)}\right).
\label{eq:magop}
\end{align}
The indices of symmetric tensor $d_{pq3}$ run over
$p=4,\cdots,\,7$. $\hat{J_3}$ and $\hat{J}_{p}$ denote the third
and the $p$th components of the spin operator acting on the soliton. 
$D_{ab}^{(\nu)} (R)$ stand for the SU(3) Wigner matrices in the
representation $\nu$, which arise from the quantization of the
soliton. $D_{\mathcal{Q}3}^{(8)}$ is defined by the combination of the SU(3)
Wigner $D$ functions
\begin{align}
D_{\mathcal{Q}3}^{(8)} = \frac12 \left( D_{33}^{(8)} + \frac1{\sqrt{3}}
  D_{83}^{(8)}\right),
\end{align}
which is obtained from the SU(3) rotation of the electromagnetic octet 
current. The coefficients $w_i$ in Eq.~\eqref{eq:magop} encode a
concrete dynamics of the chiral soliton and are independent of baryons 
involved. In fact, $w_1$ includes the leading-order contribution, a
part of the rotational $1/N_c$ corrections, and linear
$m_{\mathrm{s}}$ corrections, whereas $w_2$ and $w_3$ represent the
rest of the rotational $1/N_c$ corrections. The $m_{\mathrm{s}}$
dependent term in $w_1$ is not explicitly involved in the breaking of
flavor SU(3) symmetry. Thus, we will treat $w_1$ as if it had
contained the SU(3) symmetric part, when the magnetic moments are
computed. On the other hand, $w_4$, $w_5$, and $w_6$ are indeed the
SU(3) symmetry breaking terms. Yet another $m_{\mathrm{s}}$
corrections will come from the collective wave functions, which we
will discuss soon. In principle, $w_i$ can be computed within a
specific chiral solitonic model such as the $\chi$QSM~\cite{Kim:1995mr,
Kim:1995ha}. 

We want to emphasize that the structure of Eq.~(\ref{eq:magop}) is
rather \emph{model-independent} and is deeply rooted in the hedgehog
Ansatz or hedgehog symmetry. Since we consider the embedding of the
SU(2) soliton into SU(3)~\cite{Witten:1983tx}, which keeps the
hedgehog symmetry preserved, we have $\mathrm{SU(2)}_T \times
\mathrm{U(1)}_Y$ symmetry. So, the structure of the collective
operator is determined by the $\mathrm{SU(2)}_T \times
\mathrm{U(1)}_Y$ invariant tensors  
\begin{align}
d_{abc} = \frac14 \mathrm{tr}(\lambda_a\{\lambda_b,\,\lambda_c\}),\;\;\;
S_{ab3} = \sqrt{\frac13} (\delta_{a3}\delta_{b8} + \delta_{b3}
  \delta_{a8}),\;\;\; 
F_{ab3} = \sqrt{\frac13} (\delta_{a3}\delta_{b8} - \delta_{b3} \delta_{a8}).
\end{align}
In this respect, we will determine $w_i$ by using the experimental
data on the magnetic moments of the baryon octet as done in
Refs.~\cite{Kim:1997ip, Yang:2015era, Kim:2005gz}, instead of relying
on a specific model. We will briefly show how to fix $w_i$, using the
experimental data in the next Section.

To obtain the magnetic moments of the heavy baryons, the operator
$\hat{\mu}$ in Eq.~\eqref{eq:MagMomOp} needs to be sandwiched between
heavy baryon states. Since we consider the linear $m_s$ corrections
perturbatively, the collective wave functions for the soliton
consisting of the light-quark pair are no longer pure states. The
collective Hamiltonian for flavor SU(3) symmetry
breaking~\cite{Yang:2016qdz}, which is expressed as  
\begin{align}
  \label{eq:Hamiltonian}
  H_{\mathrm{sb}} = \alpha D_{88}^{(8)} + \beta \hat{Y} +
  \frac{\gamma}{\sqrt{3}} \sum_{i=1}^3 D_{8i}^{(8)} \hat{J}_i,
\end{align}
brings about the mixing of the baryon wave functions with those in
higher SU(3) representations. The parameters $\alpha$, $\beta$, and
$\gamma$ for heavy baryons are written as
\begin{align}
\alpha=\left (-\frac{\overline{\Sigma}_{\pi N}}{3m_0}+\frac{
  K_{2}}{I_{2}}\overline{Y}  
\right )m_{\mathrm{s}},
 \;\;\;  \beta=-\frac{ K_{2}}{I_{2}}m_{\mathrm{s}}, 
\;\;\;  \gamma=2\left ( \frac{K_{1}}{I_{1}}-\frac{K_{2}}{I_{2}} 
 \right ) m_{\mathrm{s}}.
\label{eq:alphaetc}  
\end{align}
Note that the three parameters $\alpha$, $\beta$, and $\gamma$ are 
expressed in terms of the moments of inertia $I_{1,\,2}$ and
$K_{1,\,2}$. The valence parts of them are  different from those in
the light baryon sector by the color factor $N_c-1$ in place of
$N_c$. The expression of $\overline{\Sigma}_{\pi N}$ is similar to the
$\pi N$ sigma term again except for the $N_c$ factor:
$\overline{\Sigma}_{\pi N} = (N_c-1)N_c^{-1} \Sigma_{\pi N}$.  
As mentioned previously, a singly heavy baryon consists of $N_c-1$
light valence quarks, so the constraint imposed on the right
hypercharge should be changed from $\overline{Y}=-Y'=N_c/3$ to 
$\overline{Y}=(N_c-1)/3$.  

Then the wave functions for the baryon anti-triplet ($J=0$) and
the sextet ($J=1$) are obtained respectively as~\cite{Kim:2018}   
\begin{align}
&|B_{\overline{\bm3}_{0}}\rangle = |\overline{\bm3}_{0},B\rangle + 
p^{B}_{\overline{15}}|\overline{\bm{15}}_{0},B\rangle, \cr
&|B_{\bm6_{1}}\rangle = |{\bm6}_{1},B\rangle +
  q^{B}_{\overline{15}}|{\overline{\bm{15}}}_{1},B 
\rangle + q^{B}_{\overline{24}}|{
{\overline{\bm{24}}}_{1}},B\rangle,
\label{eq:mixedWF1}
\end{align}
with the mixing coefficients
\begin{eqnarray}
p_{\overline{15}}^{B}
\;\;=\;\;
p_{\overline{15}}\left[\begin{array}{c}
-\sqrt{15}/10\\
-3\sqrt{5}/20
\end{array}\right], 
& 
q_{\overline{15}}^{B}
\;\;=\;\;
q_{\overline{15}}\left[\begin{array}{c}
\sqrt{5}/5\\
\sqrt{30}/20\\
0
\end{array}\right], 
& 
q_{\overline{24}}^{B}
\;\;=\;\;
q_{\overline{24}}\left[\begin{array}{c}
-\sqrt{10}/10\\
-\sqrt{15}/10\\
-\sqrt{15}/10
\end{array}\right],
\label{eq:pqmix}
\end{eqnarray}
respectively, in the basis $\left[\Lambda_{Q},\;\Xi_{Q}\right]$ for
the anti-triplet and $\left[\Sigma_{Q}\left(\Sigma_{Q}^{\ast}\right),\;
  \Xi_{Q}^{\prime}\left(\Xi_{Q}^{\ast}\right),\;\Omega_{Q}
  \left(\Omega_{Q}^{\ast}\right)\right]$ for the sextets. The
parameters $p_{\overline{15}}$, $q_{\overline{15}}$, and
$q_{\overline{24}}$ are given by 
\begin{eqnarray}
p_{\overline{15}}
\;\;=\;\;
\frac{3}{4\sqrt{3}}\alpha\overline{I}_{2}, 
& 
q_{\overline{15}}
\;\;=\;\;
{\displaystyle -\frac{1}{\sqrt{2}}
\left(\alpha+\frac{2}{3}\gamma\right)
\overline{I}_{2}}, 
& 
q_{\overline{24}}\;\;=\;\;
\frac{4}{5\sqrt{10}}
\left(\alpha-\frac{1}{3}\gamma\right)
\overline{I}_{2},
\label{eq:pqmix2}
\end{eqnarray}
where $\overline{I}_{2}=(N_{c}-1) N_{c}^{-1} I_{2}$. 

To carry out actual computation, we need to know the explicit
expression of the wave functions formally given in
Eq.~\eqref{eq:mixedWF1}. The wave function of a state with flavor
$F=(Y,T,T_3)$ and spin $S=(Y'=-2/3,J,J_3)$ in the representation $\nu$
is obtained in terms of a tensor with two indices, i.e. $\psi_{(\nu;\,
  F),(\overline{\nu};\,\overline{S})}$, one running over the states $F$ in the
representation $\nu$ and the other one over the states $\overline{S}$
in the representation $\overline{\nu}$. Here, $\overline{\nu}$ stands
for the complex conjugate of the representation $\nu$, and the complex
conjugate of $S$ is written as $\overline{S}=(2/3,\,J,\,J_3)$.  $N_c$
being taken to be 3, the spin with hypercharge is given by
$\overline{S}=(2/3,\,J,\,J_3)$. Thus, the collective wave function for
the soliton with a light-quark pair is expressed as 
\begin{align}
  \label{eq:SolitonWF1}
\psi_{(\nu;\, F),(\overline{\nu};\,\overline{S})}(R) =
  \sqrt{\mathrm{dim}(\nu)} (-1)^{\mathcal{Q}_S} [D_{F\,S}^{(\nu)}(R)]^*,
\end{align}
where $\mathrm{dim}(\nu)$ denotes the dimension of the representation
$\nu$ and $\mathcal{Q}_S$ a charge corresponding to the baryon state $S$,
i.e. $\mathcal{Q}_S=J_3+Y'/2$.  

To construct the complete wave function for a heavy baryon, we need to
couple the soliton wave function to the heavy quark such that the
heavy baryon becomes a color singlet. Thus, the wave
function for the heavy baryon should be written as
\begin{align}
\Psi_{B_{Q}}^{(\mathcal{R})}(R)
 =  \sum_{J_3,\,J_{Q3}} 
C_{\,J,J_3\, J_{Q}\,J_{Q3}}^{J'\,J_{3}'}
\;\mathbf{\chi}_{J_{Q3}}
\;\psi_{(\nu;\,Y,\,T,\,T_{3})(\overline{\nu};\,Y^{\prime},\,J,\,J_3)}(R)
\label{eq:HeavyWF}
\end{align}
where $\chi_{J_{Q3}}$ denote the Pauli spinors and $C_{\,J,J_3\,
  J_{Q}\,J_{Q3}}^{J'\,J_{3}'}$ the Clebsch-Gordan
coefficients. Using the wave functions given in
Eq.~\eqref{eq:HeavyWF}, we can derive the magnetic moments of the
heavy baryons  
\begin{equation}
\mu_{B}=\mu_{B}^{(0)}+\mu_{B}^{(\mathrm{op})}+\mu_{B}^{(\mathrm{wf})}
\label{eq:mu_B}
\end{equation}
where $\mu_{B}^{(0)}$ represents the part of the magnetic moment in
the chiral limit and $\mu_{B}^{(\mathrm{op})}$ arises from
$\hat{\mu}^{(1)}$ in Eq.~\eqref{eq:MagMomOp}, which contain $w_4$,
$w_5$, and $w_6$. $\mu_{B}^{(\mathrm{wf})}$ comes from the
interference between the $\mathcal{O}(m_{\mathrm{s}})$ and
$\mathcal{O}(1)$ parts of the collective wave functions given in
Eq.~\eqref{eq:mixedWF1}.   

\section{Magnetic moments of the baryon sextet}
In the present approach, it is trivial to compute the magnetic moments
of the baryon anti-triplet that consists of the soliton with spin
$J=0$. Since any scalar particle does not carry the magnetic moment,
the magnetic moments of the baryon anti-triplet all turn out to
vanish~\cite{Savage:1994zw, Faessler:2006ft}. This implies that the
magnetic moments of $\Lambda_c^+$ and $\Xi_c$ should be tiny. So, we
concentrate in this work on the magnetic moments of the baryon sextet.  

We start with the magnetic moments in the chiral limit, for which we
need to consider $\mu_B^{(0)}$ in Eq.~\eqref{eq:mu_B}. As mentioned
already, $w_1$ contains both the leading-order and a part of the
$1/N_c$ rotational corrections. Since we will use the experimental
data to fit $w_1$, it is difficult to decompose these two different
terms as already discussed in Ref.~\cite{Kim:2017khv}. Thus, we will
follow the argument of Ref.~\cite{Kim:2017khv} to fit the parameters
required for the magnetic moments of the baryon sextet. 
Using the explicit expression of the magnetic moments given
in Refs.~\cite{Kim:1995mr, Praszalowicz:1998j}, we can write  $w_1$,
$w_2$, and $w_3$ as 
\begin{align}
  \label{eq:w123}
w_{1}  =  
M_{0}\;-\;\frac{M_{1}^{\left(-\right)}}{I_{1}^{\left(+\right)}},\;\;\;
w_{2}  = -2\frac{M_{2}^{\left(-\right)}}{I_{2}^{\left(+\right)}},\;\;\;
w_{3}  = -2\frac{M_{1}^{\left(+\right)}}{I_{1}^{\left(+\right)}},
\end{align}
where the explicit forms of $M_0$, $M_1^{(\pm)}$, $M_2^{(-)}$ can be
found in Refs.~\cite{Kim:1995mr, Praszalowicz:1998j}. $I_1^{(+)}$ and
$I_2^{(+)}$ denote the moments of inertia with the notation of
Ref.~\cite{Praszalowicz:1998j} taken. It was shown in
Ref.~\cite{Praszalowicz:1998j} that in the limit of the small soliton
size the parameters in Eq.~\eqref{eq:w123} can be expressed as  
\begin{align}
M_{0}\;\rightarrow\;-2N_{c}K,
\;\;\;
\frac{M_{1}^{\left(-\right)}}{I_{1}^{\left(+\right)}}
\;\rightarrow\;\frac{4}{3}K, \;\;\;
   \frac{M_{1}^{\left(+\right)}}{I_{1}^{\left(+\right)}}
  \;\rightarrow\;-\frac{2}{3}K,\;\;\;
\frac{M_{2}^{\left(-\right)}}{I_{2}^{\left(+\right)}}
\;\rightarrow\;-\frac{4}{3}K.
\label{eq:sss}  
\end{align} 
These results with the small soliton size lead to the expressions of
the magnetic moments in the nonrelativistic (NR) quark model. 
Indeed, Eq.~\eqref{eq:w123} reproduces the correct ratio of the proton
and magnetic moments $\mu_p/\mu_n=-3/2$. So, the limit of the small
soliton size is identical to the NR limit~\cite{Praszalowicz:1998j}.   
In the NR limit, we also obtain the relation
$M_{1}^{\left(-\right)}\;=\;-2M_{1}^{\left(+\right)}$. To carry on the
computation, we have to assume that this relation is also valid in the
case of the realistic soliton size. Then, we are able to express the 
leading-order contribution $M_0$ in terms of $w_1$ and $w_3$
\begin{align}
  \label{eq:4}
M_0= w_1 + w_3.  
\end{align}
Since a heavy baryon consists of $N_c-1$ valence quarks, the original
$M_0$ should be modified by introducing $(N_c-1)/N_c$ as done
similarly in Ref.~\cite{Kim:2017khv}. The denominator $N_c$ will
cancel the same $N_c$ factor in the original expressions of $w_1$ and
$w_3$ (for explicit expressions of $w_1$ and $w_3$, we refer to 
Ref.~\cite{Kim:2018nqf}) so that they have the proper prefactor
$N_c-1$ arising from the presence of the $N_c-1$ valence quarks inside
a heavy baryon. Theoretically, only the valence part of $M_0$
requires this scaling factor. However, as far as we fix the values of
$w_i$ using the experimental data, it is not possible to fix
separately the valence and sea parts. Thus, it is plausible to define
$\tilde{w}_1$ as    
\begin{align}
\label{eq:w1tilde}
\tilde{w}_1 = \left[\frac{N_c-1}{N_c} (w_1+w_3) - w_3\right] \sigma,
\end{align}
where $\sigma$ is introduced to compensate possible deviations arising
from the relation $M_{1}^{\left(-\right)}\;=\;-2M_{1}^{\left(+\right)}$ assumed to be
valid in the realistic soliton case, and from the scaling factor
$(N_c-1)/N_c$ introduced to $M_0$ without separation of the valence
and sea parts. We want to mention that $\sigma$ has been already
determined in Ref.~~\cite{Kim:2017khv}: $\sigma\sim0.85$.

While $w_2$ and $w_3$ are kept to be the same as in the case of light
baryons, $w_{4,5,6}$ are required to be modified by introducing the
same factor $\left(N_{c}-1\right)/N_{c}$ as done for $M_0$. So,
we redefine $\overline{w}_{4,\,5,\,6}$ as 
\begin{align}
  \label{eq:tildeW}
\overline{w}_i = \frac{(N_c-1)}{N_c} w_i,\;\;\;i=4,\,5,\,6.
\end{align}
Employing the numerical values provided in Ref.~\cite{Yang:2015era}
and Eq.~\eqref{eq:tildeW}, we obtain the following values
\begin{align}
\tilde{w_{1}} 
& =  
-10.08\pm0.24,
\cr
w_{2} & =  4.15\pm0.93,
\cr
w_{3} & =  8.54\pm0.86,
\cr
\overline{w}_{4} & =  -2.53\pm0.14,
\cr
\overline{w}_{5} & =  -3.29\pm0.57,
\cr
\overline{w}_{6} & =  -1.34\pm0.56.
\label{eq:numW}
\end{align}
Using these numerical values, we can now derive the magnetic moments
of the baryon sextet. So, we want to emphasize that there is no
additional free parameter to fit.   

The explicit expressions of the magnetic moments for the baryon sextet
with $J'=1/2$ in
Eq.~\eqref{eq:mu_B} are derived as 
\begin{align}
\mu^{\left(0\right)}\left[6_{1}^{1/2},\;B_{c}\right] 
& =  
-\frac{1}{30}
\left(3\mathcal{Q}-2\right)
\left(\tilde{w}_{1}-\frac{1}{2}w_{2}-\frac{1}{3}w_{3}\right),
\label{eq:61sym}\\
\mu^{\left(\mathrm{op}\right)}
\left[6_{1}^{1/2},\;B_{c}\right] 
& =  
-\frac{1}{270}
\left[\begin{array}{c}
5\mathcal{Q}-7\\
7\mathcal{Q}-2\\
\mathcal{Q}+3
\end{array}\right]\overline{w}_{4}
\;-\;\frac{1}{90}
\left[\begin{array}{c}
4\mathcal{Q}-5\\
2\mathcal{Q}-1\\
\mathcal{Q}+3
\end{array}\right]\overline{w}_{5},
\label{eq:61opc}\\
\mu^{\left(\mathrm{wf}\right)}\left[6_{1}^{1/2},\;B_{c}\right] 
& =  -\frac{1}{90\sqrt{2}}
\left[\begin{array}{c}
4\left(\mathcal{Q}-2\right)\\
5\mathcal{Q}-4\\
\mathcal{Q}
\end{array}\right]
\left(\tilde{w}_{1}+\frac{1}{2}w_{2}+w_{3}\right)q_{\overline{15}}
\cr
 &   
\;+\;\frac{1}{90\sqrt{10}}
\left[\begin{array}{c}
1\\
2\\
3
\end{array}\right]
\left(\mathcal{Q}+1\right)
\left(\tilde{w}_{1}+2w_{2}-2w_{3}\right)q_{\overline{24}},
\label{eq:61wfc}
\end{align}
in the basis of $\left[\Sigma_{c},\,\Xi_{c}^{\prime},\,\Omega_{c}\right]$.
$\mathcal{Q}$ denotes the electric charge of the corresponding
baryon. For the baryon sextet with spin $J'=3/2$, we obtain 
\begin{align}
\mu^{\left(0\right)}\left[6_{1}^{3/2},\;B_{c}\right] 
& =  
-\frac{1}{20}\left(3\mathcal{Q}-2\right)
\left(\tilde{w}_{1}-\frac{1}{2}w_{2}-\frac{1}{3}w_{3}\right),
\label{eq:63sym}\\
\mu^{\left(\mathrm{op}\right)}
\left[6_{1}^{3/2},\;B_{c}\right] 
& =  -\frac{1}{180}
\left[\begin{array}{c}
5\mathcal{Q}-7\\
7\mathcal{Q}-2\\
\mathcal{Q}+3
\end{array}\right]\overline{w}_{4}
\;-\;\frac{1}{60}
\left[\begin{array}{c}
4\mathcal{Q}-5\\
2\mathcal{Q}-1\\
\mathcal{Q}+3
\end{array}\right]\overline{w}_{5},
\label{eq:63opc}\\
\mu^{\left(\mathrm{wf}\right)}
\left[6_{1}^{3/2},\;B_{c}\right] 
& =  -\frac{1}{60\sqrt{2}}
\left[\begin{array}{c}
4\left(\mathcal{Q}-2\right)\\
5\mathcal{Q}-4\\
\mathcal{Q}
\end{array}\right]
\left(\tilde{w}_{1}+\frac{1}{2}w_{2}+w_{3}\right)q_{\overline{15}}
\cr
 &   \;+\;\frac{1}{60\sqrt{10}}
\left[\begin{array}{c}
1\\
2\\
3
\end{array}\right]
\left(\mathcal{Q}+1\right)
\left(\tilde{w}_{1}+2w_{2}-2w_{3}\right)q_{\overline{24}},
\label{eq:63wfc}
\end{align}
in the basis of 
$\left[\Sigma_{c}^{\ast},\,\Xi_{c}^{\ast},\,\Omega_{c}^{\ast}\right]$ for
the charmed sextet.

Before we present the numerical results of the magnetic moments of the
heavy baryons, we first discuss general relations we find in
Eqs.~\eqref{eq:63opc} and \eqref{eq:63wfc}. In the present mean-field
approach, there is no difference between charmed and bottom
baryons, since there is no contribution from the heavy quark in the
limit of $m_Q\to \infty$. Thus, even though the electric charges of
the charm and bottom baryons are different each other, we have exactly
the same numerical values of the magnetic moments for both the charm
and bottom baryon belonging to the same representation, which 
can be written as 
\begin{align}
\mu\left[\mathcal{R}^{J},\;B_{c}\right]  = 
\mu\left[\mathcal{R}^{J},\;B_{b} \right].
\label{eq:mbc}
\end{align}
Furthermore, we find a general interesting relations between the
magnetic moments of the baryon sextet with $J'=1/2$ and those with
$J'=3/2$. The difference is just fact $3/2$ given as 
\begin{eqnarray}
\mu\left[6_{1}^{3/2},\;B_{c}\right] 
& = & 
{\displaystyle \frac{3}{2}}\;\mu\left[6_{1}^{1/2},\;B_{c}\right].
\label{eq:ratio}
\end{eqnarray}
In fact, this relation was already found in both the bound-state
approach of the Skyrme model~\cite{Oh:1991ws} and the SU(3) quark
models~\cite{Johnson:1976mv,Choudhury,Lichtenberg:1976fi}.  

Coleman and Glashow found various relations between the magnetic
moments of the baryon octet~\cite{Coleman:1961jn}, which arise from
the isospin invariance. Similar relations have been also obtained in
the case of the baryon decuplet~\cite{Jenkins:1994md, Kim:1997ip}.
We find here the generalized Coleman-Glashow relations for the spin-$1/2$
baryon sextet as 
\begin{align}
\mu(\Sigma_{c}^{++})\;-\;\mu(\Sigma_{c}^{+})
& =  
\mu(\Sigma_{c}^{+})\;-\;\mu(\Sigma_{c}^{0}),
\cr
\mu(\Sigma_{c}^{0})\;-\;\mu(\Xi_{c}^{\prime0}) 
& =  
\mu(\Xi_{c}^{\prime0})\;-\;\mu(\Omega_{c}^{0}),
\cr
2 [\mu(\Sigma_{c}^{+})\,-\,\mu(\Xi_{c}^{\prime0})]
& =  
\mu(\Sigma_{c}^{++})\,-\,\mu(\Omega_{c}^{0}).
\label{eq:coleman}  
\end{align}
Similar relations were also discussed in Ref.~\cite{Banuls:1999mu}.
Though the usual Coleman-Glashow relations are satisfied in the chiral
limit, the relations in Eq.~\eqref{eq:coleman} are the robust ones
even when the effects of SU(3) flavor symmetry breaking are taken into
account. In the chiral limit, we obtain the relation according to the
$U$-spin symmetry
\begin{align}
\mu(\Sigma_{c}^{0})\;=\;
\mu(\Xi_{c}^{\prime0})\;=\;
\mu(\Omega_{c}^{0})\;=\;
-2\mu(\Sigma_{c}^{+})\;=\;
-2\mu(\Xi_{c}^{\prime+})\;=\;
-\frac{1}{2}\mu (\Omega_{c}^{0}).
\label{eq:Usym}
\end{align}
Another interesting relation in the chiral limit is the sum rule given
as    
\begin{align}
\sum_{B_c\in\mathrm{sextet}}\mu(B_c)\;=\;0.
\label{eq:sum}
\end{align}
We know that Eq.~\eqref{eq:sum} is very similar to the sum rule for
the magnetic moments of the baryon decuplet.  In the case of the
decuplet, the sum of all the magnetic moments is the same as that of
all the electric charges of the corresponding
baryons~\cite{Kim:1997ip}. On the other hand, Eq.~\eqref{eq:sum} is
identical to the sum of $2\mathcal{Q}-1$ for all the members of the
baryon sextet as shown in Eq.~\eqref{eq:63sym}, which yields also the
null result as given in Eq.~\eqref{eq:sum}. 
We can derive the same relations from
Eqs.~(\ref{eq:coleman}-\ref{eq:sum}) for the $J'=3/2$ sextet baryon
and also for the bottom baryons.  

Concerning the magnetic moments of the baryon anti-triplet, we already
mentioned that those of all members turn out to be zero, since the
spin of the soliton with a light-quark pair is zero. Lichtenberg
derived a relation based on the quark model, which shows that all the
magnetic moments of the baryon anti-triplet are the
same~\cite{Lichtenberg:1976fi}  
\begin{align}
  \label{eq:2}
\mu(\Lambda_c^+) =  \mu(\Xi_c^+) = \mu(\Xi_c^0). 
\end{align}
This relation is trivially satisfied because all of them vanish in the
present work.  

\section{Numerical results}
\begin{table}[htp]
\caption{Numerical results of the magnetic moments for the charmed
  baryon sextet with $J'=1/2$ in units of the nuclear magneton $\mu_N$.} 
\renewcommand{\arraystretch}{1.3}
\begin{tabular}{ccccccc}
\hline \hline
$\mu\left[6_{1}^{1/2},\;B_{c}\right]$ 
& $\mu^{(0)}$ 
& $\mu^{(\text{total})}$ 
& Oh et al. \cite{Oh:1991ws} 
& Scholl and Weigel \cite{Scholl:2003ip} 
& Faessler et al. \cite{Faessler:2006ft}
& Lattice QCD~\cite{Can:2013tna, Bahtiyar:2016dom} 
\tabularnewline \hline
$\Sigma_{c}^{\text{++}}$ 
& $2.00\pm0.09$
& $2.15\pm0.1$ 
& $1.95$
& $2.45$ 
& $1.76$
& $2.220\pm 0.505$
\tabularnewline
$\Sigma_{c}^{\text{+}}$ 
& $0.50\pm0.02$ 
& $0.46\pm0.03$ 
& $0.41$ 
& $0.25$ 
& $0.36$
& --
\tabularnewline
$\Sigma_{c}^{0}$ 
& -$1.00\pm0.05$ 
& -$1.24\pm0.05$ 
& -$1.1$ 
& -$1.96$ 
& -$1.04$
& -$1.073\pm 0.269$
\tabularnewline
\hline 
$\Xi_{c}^{\prime+}$ 
& $0.50\pm0.02$ 
& $0.60\pm0.02$ 
& $0.77$ 
& --
& $0.47$ 
& $0.315\pm0.141$
\tabularnewline
$\Xi_{c}^{\prime0}$ 
& -$1.00\pm0.05$ 
& -$1.05\pm0.04$ 
& -$1.12$ 
& -- 
& -$0.95$
& -$0.599\pm0.071$
\tabularnewline
\hline 
$\Omega_{c}^{0}$ 
& -$1.00\pm0.05$ 
& -$0.85\pm0.05$ 
& -$0.79$ 
& -- 
& -$0.85$
& -$0.688\pm 0.031$
\tabularnewline
\hline \hline
\end{tabular}
\label{tab:1}
\end{table}
\begin{table}[htp]
\renewcommand{\arraystretch}{1.3}
\caption{Numerical results of magnetic moments for charmed baryon
  sextet with $J'=3/2$ in units of the nuclear magneton $\mu_N$.} 
\begin{tabular}{ccccc}
\hline \hline 
$\mu\left[6_{1}^{3/2},\;B_{c}\right]$ 
& $\mu^{(0)}$ 
& $\mu^{(\text{total})}$ 
& Oh et al.\cite{Oh:1991ws}
& Lattice QCD~\cite{Can:2015exa}
\tabularnewline
\hline 
$\Sigma_{c}^{\ast\text{++}}$ 
& $3.00\pm0.14$ 
& $3.22\pm0.15$ 
& $3.23$
& --
\tabularnewline
$\Sigma_{c}^{\ast\text{+}}$ 
& $0.75\pm0.04$ 
& $0.68\pm0.04$ 
& $0.93$
&--
\tabularnewline
$\Sigma_{c}^{\ast0}$ 
& $-1.50\pm0.07$ 
& $-1.86\pm0.07$ 
& $-1.36$
&--
\tabularnewline
\hline 
$\Xi_{c}^{\ast+}$ 
& $0.75\pm0.04$ 
& $0.90\pm0.04$ 
& $1.46$
&--
\tabularnewline
$\Xi_{c}^{\ast0}$ 
& $-1.50\pm0.07$ 
& $-1.57\pm0.06$ 
& $-1.4$
&--
\tabularnewline
\hline 
$\Omega_{c}^{\ast0}$ 
& -$1.50\pm0.07$ 
& -$1.28\pm0.08$ 
& -$0.87$
& -$0.730\pm0.023$
\tabularnewline
\hline \hline
\end{tabular}
\label{tab:2}
\end{table}
In Table~\ref{tab:1} and~\ref{tab:2}, we list the results of the
magnetic moments for the charmed baryon sextet with $J'=1/2$ and
$J'=3/2$, respectively. In the second columns, the 
the numerical values of the magnetic moments in the chiral limit are
listed whereas in the third columns the total results are given. The
effects of flavor SU(3) symmetry breaking contribute to the magnetic
moments of the baryon sextet found to be around
$(5-10)\,\%$ except for those of $\Sigma_c^0$ and $\Omega_c^0$, for
which the effects are found to be around $(15-20)\,\%$. 

In Table~\ref{tab:1}, we compare the results for the baryon sextet
with $J'=1/2$ with those from the Skyrme model in the
bound-state approach~\cite{Oh:1991ws, Scholl:2003ip}, the
relativistic quark model~\cite{Faessler:2006ft}, and the lattice
QCD~\cite{Can:2013tna, Can:2015exa, Bahtiyar:2016dom}.  
While the original Skyrme model in the bound-state
approach~\cite{Oh:1991ws} is constructed, based on the light
pseudoscalar meson fields together with the heavy pseudoscalar meson
fields, Ref.~\cite{Scholl:2003ip} introduced the light and heavy vector mesons
in addition. We find that in general the magnitudes of the present
results are slightly larger than those of Ref.~\cite{Oh:1991ws} except
for $\mu(\Xi_c^{+})$ and $\mu(\Xi_c'^{0})$. The results of
$\mu(\Sigma_c^{++})$ and $\mu(\Sigma_c^0)$ from
Ref.~\cite{Scholl:2003ip} are, respectively, around $15\,\%$ and
$60\,\%$ larger than the present results in magnitude. 
The results are qualitatively similar to those of
Refs.~\cite{Faessler:2006ft}, though Ref.~\cite{Faessler:2006ft} used
a rather different model, i.e. the relativistic quark model.
Comparing the present results with those from the lattice
QCD~\cite{Can:2013tna, Can:2015exa, Bahtiyar:2016dom}, we find that
the results are systematically larger than those from the lattice QCD
except for the $\Sigma_c^{++}$ magnetic moment.  

Since the results of the magnetic moments of the bottom baryons are
exactly the same as those of the charmed baryons, it seems redundant
to present the corresponding results. The present results for the
magnetic moments of the bottom baryon sextet are even in better
agreement with those from Ref.~\cite{Faessler:2006ft}. 

\section{Summary and conclusion}
In the present work, we have studied the magnetic moments of the 
lowest-lying singly heavy baryons, based on the chiral quark-soliton
model. All the dynamical parameters were fixed in the light baryon 
sector. The magnetic moments of the baryon anti-triplet vanish in the
present mean-field approach, because the spin of the soliton for the
anti-triplet is $J=0$. Since the first term in the expression of the
heavy baryon magnetic moments consists of the leading-order and the
$1/N_c$ rotational corrections, we had to decompose them, using the
limit of the small soliton size. Having properly considered the
scaling factor, we were able to compute the magnetic moments of the
baryon sextet with both spins $J'=1/2$ and $J'=3/2$. The results were 
compared with those from other models such as the Skyrme model in
bound-state approaches, the relativistic quark model, and the lattice
QCD. They are in particular consistent with those from the
relativistic quark model. We also compared the present results with
those from the lattice QCD.  Except for the $\Sigma_c^{++}$ magnetic
moment, we obtained systematically larger values of the magnetic
moments than those from the lattice QCD. 

The same method can be applied to compute the transition magnetic
moments of the singly heavy baryons, which provide essential
information on radiative decays of them. The corresponding
investigation is under way. The magnetic moments of doubly heavy
baryons can be also studied within the present mean-field approach,
assuming that $N_c-2$ valence quarks can produce the modified pion
mean fields. The corresponding study is under investigation. 

\section*{Acknowledgments} 
The authors are grateful to M. V. Polyakov and M. Prasza{\l}owicz for
valuable discussion and suggestions in the early stage of the present
work. H.-Ch. K is thankful to A. Hosaka and T. Nakano at the Nuclear
Physics (RCNP), Osaka University for hospitality, where part of the
work was done. The present work was supported by Basic Science
Research Program through the National Research Foundation of Korea
funded by the Ministry of Education, Science and Technology (Grant 
No. NRF-2015R1D1A1A01060707(H.-Ch.K.) and  Grant
No. NRF-2016R1C1B1012429 (G.-S. Y.)).  

\end{document}